\documentclass[twocolumn,epjc3]{svjour3} 

\RequirePackage[T1]{fontenc}

\smartqed 

\RequirePackage[numbers,sort&compress]{natbib}
\RequirePackage[colorlinks,citecolor=blue,urlcolor=blue,linkcolor=blue]{hyperref}
\RequirePackage{flushend}

\RequirePackage[dvipsnames]{xcolor}
\RequirePackage{tikz}
\RequirePackage[compat=1.1.0]{tikz-feynhand}
\usetikzlibrary{decorations.pathmorphing}
\usetikzlibrary{decorations.markings}

\RequirePackage{amsmath}
\RequirePackage{amssymb}
\RequirePackage{graphicx}
\RequirePackage{epsfig}
\RequirePackage{mathptmx}  

\DeclareMathOperator{\Df}{{\mathfrak{ D}}}
\DeclareMathOperator{\DDf}{{\mathcal{ D}}}
\DeclareMathOperator{\bx}{{\bf{ x}}}

\journalname{Eur. Phys. J. C}

\begin{document}

\title{Semiclassical Approximation meets Keldysh-Schwinger diagrammatic technique: Scalar $\varphi^4$  }

\author{A.A. Radovskaya \thanksref{e1}
 \and 
        A.G. Semenov\thanksref{e2}
}

\thankstext{e1}{e-mail: raan@lpi.ru}
\thankstext{e2}{e-mail: semenov@lpi.ru}

\institute{ P.N.Lebedev Physical Institute, 119991, Moscow, Russia}

\date{Received: date / Accepted: date}

\maketitle

\begin{abstract}
We study the evolution of the non-equilibrium quantum fields from a highly excited initial state in two approaches: 
the standard Keldysh-Schwinger diagram technique and the semiclassical expansion.
We demonstrate explicitly that these two approaches coincide if the coupling constant $g$  and the Plank constant $\hbar$ are small simultaneously.
Also, we discuss loop diagrams of the perturbative approach, which are summed up by the leading order term of the semiclassical expansion. 
 As an example, we consider shear viscosity for the scalar field theory at the leading semiclassical order. 
 We  introduce the new technique that unifies both semiclassical and diagrammatic approaches and 
 open the possibility to perform the resummation of the semiclassical contributions.
 \end{abstract}

\section{Introduction}
\label{sec:intro}
Highly nonequilibrium dense quantum fields define the initial stage of many physical problems.

These include physics of the early stage of ultrarelativistic heavy ion collisions \cite{G2015,K2016,berg1}, cold atomic gases \cite{berges_gases,Lee} and the processes in the early Universe \cite{cosm_1,cosm_2,cosm_3,akh2016,akh2018}. At present, there is a variety of approaches, which is used for the description of the quantum field evolution from a highly excited initial state to the quasistationary one where hydrodynamic equations work well.

One of the most advanced approaches is the Keldysh-Schwinger diagram technique which provides a systematic way of studying nonequilibrium phenomena in quantum field theory \cite{kel,sch,berg1,GKSV2020}. With the help of this technique, one can derive the kinetic equations, which describe the evolution of quasiparticle distribution function and observables consequently. Also, this technique can be used for the systematic evaluation of thermodynamical and transport properties of the quantum systems at the thermal equilibrium \cite{berg1,GKSV2020}.

Another way to deal with the nonequilibrium initial state comes from the physical intuition and based on the assumption that at high energies and/or high occupation numbers the dynamics of the quantum fields is semiclassical, so one can use the classical equations of motion \cite{MM,FGM2007,initial1,initial2,initial3,LR1,LR2,LR3,method,Bodeker97,Bodeker98}. In order to complete this approach, one should make additional assumptions about ensemble which is used for the averaging of observables.  In the literature, this approach often called \emph{the Classical Statistical Approximation} (CSA). This approch is very useful for numerical simulations. It allows numerically extract (nonperturbatively in coupling constant) results for observables \cite{bogusl18,aarts01} and transport coefficients \cite{jakovac15}. In the previous works of one of the authors \cite{LR1,LR2,LR3}, it was shown that the CSA arises as the leading order of the semiclassical expansion.

This work aims to demonstrate that both the Keldysh-Schwinger diagram technique and the classical statistical approach are two
facets of one general way to deal with nonequilibrium quantum fields. 
It seems that these two approaches are quite different from a practical point of view.  The Keldysh-Schwinger diagram technique originates from the perturbative expansion in coupling constant $g$.  In order to evaluate the observable consistently in this approach, one should (with the help of the Wick theorem and the diagram technique) derive the quantum kinetic equation on the distribution function,  solve it and  evaluate observable using the solution.  On the other hand, the CSA  comes from the $\hbar$ expansion. In order to find the observable here, it is necessary to solve the classical equations of motion and to average the observable on the classical trajectories with the ensemble of the initial conditions.  In this work, we show that these two approaches can be derived from one general path-integral representation.  We demonstrate which assumptions and approximations should be made to obtain one or another approach.  We study the case of both $g$ and $\hbar$ are small, where these two approaches should be consistent, and analyse how the leading order term of the semiclassical expansion (the CSA) sums up some multi-loop diagrams of the perturbative Keldysh-Schwinger technique. 

Previously some efforts were made in this direction. In particular,  in work \cite{aarts98}, the comparison between the Keldysh-Schwinger diagram technique and the classical approximation was performed for the thermal equilibrium.  The agreement of the leading order contributions at high temperatures was demonstrated. 

The paper is organised as follows:\\
Starting from the general setup described in section \ref{sec:Keldysh}, we briefly review the standard Keldysh-Schwinger diagram technique in section \ref{sec:perturbative} and the semiclassical approximation
in section \ref{sec:semiclassical}. Then in section \ref{sec:comparison} we perform the detailed comparison of these approaches in the limit of small both $g$ and $\hbar$ and analyse loop contributions. In section \ref{sec:viscosity}, on the example of the shear viscosity, we show how to employ the CSA for calculation of the relevant physical observables and discuss the applicability of the semiclassical approach. Section \ref{sec:newdiag} is devoted to a new diagram technique which naturally combines both approaches described above. The discussion of results and conclusions are in section \ref{sec:conclusions}.

\section{Keldysh-Schwinger approach to the non-equilibrium QFT } 
\label{sec:Keldysh}

The standard way to deal with the non-equilibrium quantum field theory includes the Keldysh-Schwinger technique, also known as the closed-time path formalism \cite{berg1,GKSV2020,method}. In this approach, averages are calculated as the trace with the density matrix operator. Time evolution of the density matrix is defined by two evolution operators; that is why the doubling of the degrees of freedom occurs. Moreover,  the initial density matrix should be additionally defined from the physics of the considered system.

In this work we consider the scalar field theory with the action
\begin{multline}\label{action}
 S[\varphi(x)]=\\ 
 \frac12\int d^dx\left(\partial_{\mu}\varphi(x)\partial^{\mu}\varphi(x)-m^2\varphi^2(x)-\frac {g}{2}\varphi^4(x)\right).
\end{multline}
Here and after we use mostly minus metric convention $g_{\mu\nu} = (+,-,-,-)$ and $x^{\mu} = (t,\bx)$.
 Using eq. \eqref{action} one can calculate the Keldysh action as a difference between actions on the forward and the 
 backward parts of the Keldysh contour.
Averages are  expressed through the path integrals with the Keldysh action as  \cite{berg1,GKSV2020,method}
 \begin{gather}\label{O_phi_F}
  \langle \hat O \rangle=\int\limits \mathcal D\varphi_F\mathcal D\varphi_B O[\varphi_F,\varphi_B]
   e^{\frac{i}{\hbar}(S[\varphi_F]-S[\varphi_B])}.
 \end{gather}
 Here and after we keep $\hbar$ explicitly in order to study the semiclassical limit of the theory.
 
 It is convenient to rotate the  basis of $\varphi_F,\varphi_B$ fields to  
  so-called the "classical" and the "quantum" ones (there are equivalent notations
  for such rotation in the literature \\
   $\phi_c\equiv \phi_r$ and $\phi_q \equiv \phi_a $) \cite{berg1,GKSV2020,method}
 \begin{gather}
    \varphi_{cl}(x)=\frac12\left(\varphi_{F}(x)+\varphi_{B}(x)\right),\\ 
    \hbar\varphi_{q}(x)=\varphi_{F}(x)-\varphi_{B}(x).\nonumber
 \end{gather}

 New $\varphi_{cl},\varphi_q$ basis has few advantages: the causality of the theory become explicit, 
 the vertices look simpler, and the semiclassical limit is transparent. 
Then the Keldysh action transforms to (after integration by parts) 
\begin{gather}\label{SK}
   S[\varphi_F]-S[\varphi_B]= S_{init}[\varphi_{cl},\varphi_{q}] +S_K[\varphi_{cl},\varphi_{q}],\nonumber  \\
   S_{init}[\varphi_{cl},\varphi_{q}] = \hbar\int d^{d-1} \bx\ \varphi_q(t_0,\bx)\dot\varphi_{cl}(t_0,\bx),\nonumber \\
   S_K[\varphi_{cl},\varphi_{q}] = \\ \nonumber
    - \hbar\int\limits_{t_0}^{\infty} dt \int d^{d-1}\bx\
  \varphi_q(t,\bx)\left(\partial^2_t-\nabla^2 + m^2 \right)\varphi_{cl}(t,\bx) \nonumber
  \\-g\hbar\int\limits_{t_0}^{\infty} dt \int d^{d-1}\bx\left(\varphi_{cl}^3(t,\bx)\varphi_q(t,\bx)
  +\frac{\hbar^2}{4}\varphi_{cl}(t,\bx)\varphi_q^3(t,\bx) \right).\nonumber
\end{gather}
We keep explicit dependence on the initial time $t_0$ to take into account highly non-equilibrium initial states. 
Usually, the initial time moment is set to the past infinity, and the boundary term $ S_{init}[\varphi_{cl},\varphi_{q}]$ is dropped out.  
 Then the averages can be calculated by integration over new fields as
 \begin{gather}
  \label{O_phi_cl}
  \langle \hat O \rangle=\int \DDf \varphi_{cl}\DDf \varphi_q\ 
  O[\varphi_{cl}]\  e^{\frac{i}{\hbar}S_K[\varphi_{cl},\varphi_q]}.
 \end{gather}
 Note, if  $\hat O$ contains only equal-time operators, then it is sufficient to keep only  $\varphi_{cl}$ component 
 in the integrand of the expression eq.\eqref{O_phi_cl} due to causality.

The expressions similar to eq.\eqref{O_phi_cl} and eq.\eqref{O_phi_F} can be found in many modern textbooks discussing the Keldysh-Schwinger technique. 
However, such representation is a bit misleading:
 it does not contain information about the initial state of the theory, which makes $\varphi_F$ 
 and $\varphi_B$ fields correlated. More rigorously the eq.\eqref{O_phi_cl} can be written as \cite{LR1,LR2,LR3,zarand}
 \begin{multline}\label{common}
  \langle \hat O \rangle = 
  \int\mathfrak D\Pi(\bx)\mathfrak D\alpha(\bx)\ \mathcal W[\alpha(\bx),\Pi(\bx)]\\
  \times \int\limits_{i.c.}\mathcal
   D\varphi_{cl}(t,\bx)\int\mathcal D\varphi_q(t,\bx)  O[\varphi_{cl}]
 e^{\frac{i}{\hbar}S_K[\varphi_{cl},\varphi_q]},
\end{multline}
where the integral with $i.c.$ means the initial values for the $\varphi_{cl}$ field, 
  $\varphi_{cl}(t_0,\bx)=\alpha(\bx)$, $\partial_t\varphi_{cl}(t_0,\bx)=\Pi(\bx)$; whereas the initial values for the 
  $\varphi_q$ are  not fixed. $ S_{init}[\varphi_{cl},\varphi_{q}]$ now is taken into account and absorbed by the Wigner function.
  The Wigner function is related to the initial value of the density matrix operator $\hat\rho(t_0)$ as
\begin{multline}
  W[\alpha(\bx),\Pi(\bx)] = \int \Df \beta(\bx)  e^{i\int d^{d-1}\bx \beta(\bx)\Pi(\bx)}\\
   \times\langle  \alpha(\bx)
   + \frac{\hbar}{2}\beta(\bx)|\hat \rho(t_0)|\alpha(\bx) - \frac{\hbar}{2}\beta(\bx)\rangle.
\end{multline}
This function contains all the information about the initial state of the system. 
 The eq.\eqref{common} represents the general expression from which one can deduce both the perturbative
  and the semiclassical approaches as we discussed in the introduction. In the next section, we derive the standard
  Keldysh-Schwinger perturbation technique and discuss its limitations.

\section{Standard perturbative approach}
\label{sec:perturbative}

The standard Keldysh-Schwinger diagram technique follows naturally from two major assumptions:

- Gaussian form of the initial Wigner function that allows the Wick theorem to be valid;

- Possibility of the perturbative, in the coupling constant g, expansion.\\
Under these assumptions, the eq.\eqref{common} can be rewritten as
\begin{equation}
\langle \hat O \rangle = \left\langle O[\varphi_{cl}] 
 e^{-ig\int d^dx\left(\varphi_{cl}^3(x)\varphi_{q}(x)
 +\frac{\hbar^2}{4}\varphi_{cl}(x)\varphi_{q}^3(x)\right)} \right\rangle_0,
\end{equation}
 where the averaging over the noninteracting fields $\langle...\rangle_0$ should be performed with help of the Wick's theorem with four basic contractions \cite{berg1,GKSV2020,method}:
\begin{eqnarray}
  \langle \varphi_{cl}(x)\varphi_{cl}(x')\rangle_0&=&iG_K^0(x;x'),\nonumber\\
  \langle \varphi_{cl}(x)\varphi_{q}(x')\rangle_0&=&iG^0_R(x;x'),\nonumber\\
  \langle \varphi_{q}(x)\varphi_{cl}(x')\rangle_0&=&iG^0_A(x;x')=iG^0_R(x';x),\nonumber\\
  \langle \varphi_{q}(x)\varphi_{q}(x')\rangle_0&=&0.
\end{eqnarray}
Here $G^0_{R(A)}$ is retarded (advanced) Green functions which can be equivalently defined in the operator formalism as
\begin{equation}\label{GRA}
  G_R^0(x;x')=G_A^0(x';x)=-\frac{i}{\hbar}\theta(t-t')\langle[\hat \varphi(x),\hat \varphi(x')]\rangle_0.
\end{equation}
In the absence of the interactions these free correlators eq.\eqref{GRA} are independent from the initial Wigner function and solve the equation
\begin{gather}\label{L0}
  \hat L_0  G_{R(A)}^0(x;x')=- \delta^d(x-x'),\\
  \hat L_0 = \partial_{\mu}\partial^{\mu}+m^2
\end{gather}
with the corresponding boundary conditions (retarded or advanced ones). The solution of the eq.\eqref{L0} is, for example,
\begin{gather}
  G_{R}^0(x;x') = -\theta(t-t')\int \frac{d^{d-1}{\bf p}}{(2\pi)^{d-1}} \frac{\sin(\omega_p(t-t'))}{\omega_p} e^{-i{\bf p\cdot x}}, 
  \nonumber\\
   \omega_p^2 = {\bf p}^2 +m^2.
\end{gather}
The Keldysh Green function 
\begin{equation}
  G_K^0(x;x') = -\frac{i}{2}\langle\{\hat \varphi(x),\hat \varphi(x')\}\rangle_0
\end{equation}
solves the  equation
\begin{gather}\label{LO_GK}
  \hat L_0 G_K^0(x;x') = 0.
\end{gather}
This correlator depends crucially on the initial state of the system and can not be found without specification of the initial density operator. In the simplest case, the initial state of the system is characterised by the one-particle distribution function $f_p$, which is related to the Keldysh Green function as \cite{berg1,GKSV2020,method}
\begin{multline}\label{GK0}
  G_K^0(x;x') = -i\hbar\int \frac{d^{d-1}{\bf p}}{(2\pi)^{d-1}} 
  \frac{\cos(\omega_p(t-t'))}{4\omega_p}\\
  \times (2f_p+1) e^{-i{\bf p\cdot x}}.
\end{multline}

It is necessary to stress here that only for the Gaussian initial state the knowledge of $G_R^0(x;x')$ and $G_K^0(x;x')$  is enough to perturbatively evaluate the average of any product of the free fields and build up the diagram technique.\footnote{
  Another way to include the initial conditions is to extend the Keldysh contour 
  onto the imaginary axis to take into account the Matsubara part \cite{GKSV2020}. However, it works only for the special case of the 
  thermal initial state.}

The basic elements of each diagram are  two propagators
\begin{gather*}
  iG^0_R(x_1;x_2) \qquad \begin{tikzpicture}[baseline=(a.base)]
    \begin{feynhand}
      \vertex (a) at (0,0) {$x_2$};\vertex (b) at (3.0,0) {$x_1$};
      \propag [fer] (a) to (b);
    \end{feynhand}
  \end{tikzpicture}, \\
  iG^0_K(x_1;x_2)\qquad\begin{tikzpicture}[baseline=(a.base)]
    \begin{feynhand}
      \vertex (a) at (0,0) {$x_2$};\vertex (b) at (3.0,0) {$x_1$};
      \propag [plain] (a) to (b);
    \end{feynhand}
  \end{tikzpicture},
\end{gather*}
and two vertices
$$\begin{tikzpicture}[baseline=(v.base)]
    \begin{feynhand}
      \vertex [dot] (v) at (0,0) {};\vertex (c1) at (-0.8,0.5) {};\vertex (c2) at (-1,0) {};
      \vertex (c3) at (-0.8,-0.5) {};\vertex (q) at (1.0,0.0) {};
      \propag [plain] (c1) to (v);
      \propag [plain] (c2) to (v);
      \propag [plain] (c3) to (v);
      \propag [fer] (v) to (q);
    \end{feynhand}
  \end{tikzpicture}\quad -ig,\qquad
  \begin{tikzpicture}[baseline=(v.base)]
    \begin{feynhand}
      \vertex [ringdot] (v) at (0,0) {};\vertex (q1) at (0.8,0.5) {};\vertex (q2) at (1,0) {};
      \vertex (q3) at (0.8,-0.5) {};\vertex (c) at (-1.0,0.0) {};
      \propag [fer] (v) to (q1);
      \propag [fer] (v) to (q2);
      \propag [fer] (v) to (q3);
      \propag [plain] (c) to (v);
    \end{feynhand}
  \end{tikzpicture}\quad -\frac{ig \hbar^2}{4}.
$$
Here the "black" and the "white" vertices differ by the power of $\hbar^2$. It is specialised for the exact comparison with the semiclassical approach later.

For example, let us draw diagrams for the first two orders of the coupling constant expansion for the full
 retarded Green function  in the presence of interactions
\begin{equation}
  G_R(x,x') = -i\langle \varphi_{cl}(x)\varphi_{q}(x')\rangle.
\end{equation}

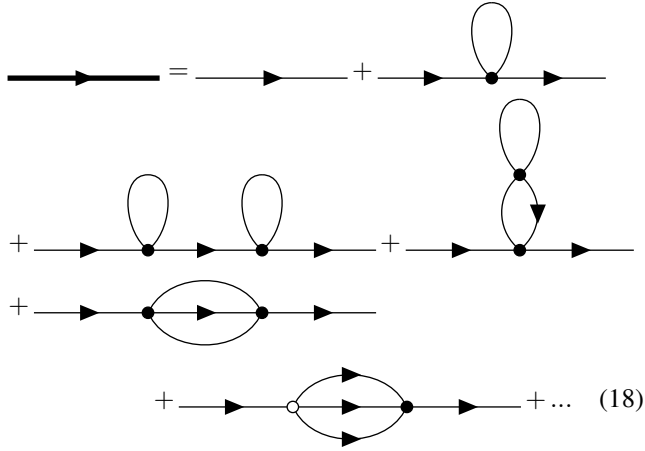
\begin{multline}\label{exp_g}
\begin{tikzpicture}[baseline=(a.base)]
    \begin{feynhand}
    \vertex (a) at (0,0);\vertex (b) at (2.0,0) ;
    \setlength{\feynhandlinesize}{2pt}
    \propag [fer] (a) to (b);
    \end{feynhand}
\end{tikzpicture}=
\begin{tikzpicture}[baseline=(a.base)]
    \begin{feynhand}
    \vertex (a) at (0,0);\vertex (b) at (2.0,0);
    \propag [fer] (a) to (b);
    \end{feynhand}
\end{tikzpicture}+
\begin{tikzpicture}[baseline=(a.base)]
    \begin{feynhand}
    \vertex (a) at (0,0);\vertex [dot] (b) at (1.5,0) {};\vertex (bb) at (1.5,1);\vertex (c) at (3.0,0);
    \propag [fer] (a) to (b);
    \propag [fer] (b) to (c);
    \propag [plain] (b) to [in=180,out=135] (bb);
    \propag [plain] (bb) to [in=45,out=0] (b);
    \end{feynhand}
\end{tikzpicture}\\
+\begin{tikzpicture}[baseline=(a.base)]
    \begin{feynhand}
    \vertex (a) at (0,0);\vertex [dot] (b) at (1.5,0) {};\vertex (bb) at (1.5,1);\vertex [dot] (c) at (3.0,0) {};\vertex (cc) at (3.0,1);\vertex (d) at (4.5,0);
    \propag [fer] (a) to (b);
    \propag [fer] (b) to (c);
    \propag [fer] (c) to (d);
    \propag [plain] (b) to [in=180,out=135] (bb);
    \propag [plain] (bb) to [in=45,out=0] (b);
    \propag [plain] (c) to [in=180,out=135] (cc);
    \propag [plain] (cc) to [in=45,out=0] (c);
    \end{feynhand}
\end{tikzpicture}   + 
\begin{tikzpicture}[baseline=(a.base)]
    \begin{feynhand}
    \vertex (a) at (0,0);\vertex [dot] (b) at (1.5,0) {};\vertex [dot] (bb) at (1.5,1) {};\vertex  (bbb) at (1.5,2);\vertex (c) at (3.0,0);
    \propag [fer] (a) to (b);
    \propag [fer] (b) to (c);
    \propag [plain] (b) to [in=225,out=135] (bb);
    \propag [fer] (bb) to [in=45,out=315] (b);
    \propag [plain] (bb) to [in=180,out=135] (bbb);
    \propag [plain] (bbb) to [in=45,out=0] (bb);
    \end{feynhand}
\end{tikzpicture}\\+
\begin{tikzpicture}[baseline=(a.base)]
    \begin{feynhand}
    \vertex (a) at (0,0);\vertex [dot] (b) at (1.5,0) {};\vertex [dot] (c) at (3.0,0) {};\vertex (d) at (4.5,0);
    \propag [fer] (a) to (b);
    \propag [fer] (b) to (c);
    \propag [fer] (c) to (d);
    \propag [plain] (b) to [in=120,out=60] (c);
    \propag [plain] (b) to [in=240,out=300] (c);
    \end{feynhand}
\end{tikzpicture}\\+
\begin{tikzpicture}[baseline=(a.base)]
    \begin{feynhand}
    \vertex (a) at (0,0);\vertex [ringdot] (b) at (1.5,0) {};\vertex [dot] (c) at (3.0,0) {};\vertex (d) at (4.5,0);
    \propag [fer] (a) to (b);
    \propag [fer] (b) to (c);
    \propag [fer] (c) to (d);
    \propag [fer] (b) to [in=120,out=60] (c);
    \propag [fer] (b) to [in=240,out=300] (c);
    \end{feynhand}
\end{tikzpicture}+...
\end{multline}
and make some important observations. The first one is related to causality. 
The zeroth-order retarded Green function $G_R^0(x;x')$ is explicitly zero if $t\leq t'$. 
It means that time increase according to arrows direction on the diagrams and each diagram
 in this expansion vanishes identically if $t\leq t'$. So, this diagrammatic expansion respects causality
  and full Green function $G_R(x;x')=0$ for $t\leq t'$ as expected. For another observation, 
  let us cut for the moment all $G_K^0$ lines and inspect what remains. One can see that the number 
  of remaining loops exactly equal to the $\hbar$ order of the diagram, i.e. twice of the number of the "white" 
  vertices. We put this observation on the solid ground below. In this place, the Keldysh-Schwinger technique differs
   drastically from the standard Feynman diagram technique, where the $\hbar$ order of any diagram coincides 
   with the number of the loops.\footnote{For the $\varphi^4$ theory considered here there is an additional
    relation between number of the loops and number of the vertices. According to this relation, 
    the number of the loops equal to the power of coupling constant. It comes from combinatorial 
    arguments and valid for the Keldysh-Schwinger
    technique considered here.}

As an example let us write the explicit expression for the "cactus"  diagram (the last diagram on the second line of eq.\eqref{exp_g})
\begin{multline}\label{cactus}
  G_R^{(cactus)}(x;x')=-18g^2\int d^dy\ d^d y'
  G^R(x;y)G^R(y;y')\\
  \times G^R(y;x')G^K(y;y')G^K(y';y').
\end{multline}
In the next section we demonstrate how this diagram (and all others) originates from the coupling constant expansion of the CSA.

\section{Semiclassical approach}
\label{sec:semiclassical}

In order to construct the semiclassical expansion, we add an auxiliary  source $J(x)$
to the theory described by eq.\eqref{action} 
\begin{multline}\label{actionJ}
  S[\varphi(x),J(x)]=\frac12\int d^dx\Big(\partial_{\mu}\varphi(x)\partial^{\mu}\varphi(x)-m^2\varphi^2(x)\\
  -\frac {g}{2}\varphi^4(x) + 2 J(x)\varphi(x)\Big).
 \end{multline}
The source $J(x)$ is used  for the intermediate steps only and should be set to zero at the and of the calculations.

Let us rewrite the Keldysh action (eq.\eqref{SK}) in a more convenient form
\begin{multline}
  S_K[\varphi_{cl},\varphi_{q},J] = - \hbar\int\limits_{t_0}^{\infty} dt \int d^{d-1}\bx\ \Big(
  \varphi_q A[\varphi_{cl}] + \frac{g\hbar^2}{4}\varphi_{cl}\varphi_q^3\Big),
  \\
  A[\varphi_{cl}] = (\partial_{\mu}\partial_{\mu}+m^2)\varphi_{cl} + g\varphi_{cl}^3 -J.
\end{multline}  
There are two  key features of this action:

 - $A[\varphi_{cl}] =0$ corresponds to projecting onto the classical equation of motion of the 
Lagrangian (eq.\eqref{actionJ}). 

 - As far as we explicitly keep $\hbar$ - dependence, it is clear that semiclassical approach is, in fact, the expansion of the last term 
 \begin{multline}\label{semi}
 e^{-i\frac{ g\hbar^2}{4}\int\limits_{t_0}^{\infty} dt \int d^{d-1}\bx\varphi_{cl}\varphi_q^3} =
  1
  -i\frac{g\hbar^2}{4}\int\limits_{t_0}^{\infty} dt \int d^{d-1}\bx\ \varphi_{cl}\varphi_q^3+\cdots
 \end{multline}

 \subsection{Classical Statistical Approximation}
 The Leading Order term of the semiclassical expansion\\
 (eq.\eqref{semi}) is also known as the Classical Statistical Approximation, or the classical approach.
 In this case, the integral over $\varphi_q$ and $\varphi_c$ fields can be done, and the eq.\eqref{common}  reproduce the well-known result \cite{MM,FGM2007,initial1,initial2,initial3,LR1,LR2,LR3}
 \begin{gather} \label{CSA}
  \langle \hat O\rangle = 
   \int \mathfrak{D}\alpha(\bx)  \mathfrak{D} 
   \Pi(\bx) W[\alpha(\bx),\Pi(\bx))] O(\phi_c),
  \end{gather}
  where
$\phi_{c}$ is the solution of the classical equation of motion
    \begin{gather}
   \partial_{\mu}\partial^{\mu}\phi_{c}+g\phi_{c}^3 = J
	 \label{classEoM}
   \end{gather}
   with initial conditions given by
   \begin{gather}
   \phi_{c}(t_0,\bx) = \alpha(\bx), \quad
   \partial_t\phi_{c}(t_0,\bx) = \Pi(\bx)
 \end{gather}
 and at zero axillary source $J(t,\bx)$.\\
 Hence, the recipe for the CSA is the following:

- find the classical trajectory as a function  of the initial conditions;

- calculate observables on this trajectory;

- average over the initial conditions with the Wigner function corresponding the considered problem.\\
Let us introduce new notation for 
 averaging over initial conditions with the Wigner function as
\begin{gather}
 \langle \cdots \rangle_{i.c.} \equiv  \int \mathfrak{D}\alpha(\bx)  \mathfrak{D} 
 \Pi(\bx) W[\alpha(\bx),\Pi(\bx))] (\cdots)
\end{gather}
Then the definition of the CSA approximation (eq. \eqref{CSA}) can be rewritten in this notation as 
\begin{gather}
  \langle \hat O \rangle =\big\langle O[\phi_c] \big\rangle_{i.c.}
\end{gather}
It may seem that in the semiclassical expansion  there are no linear in $\hbar$ contributions. 
However, it is not the case since the Wigner function may depend on the $\hbar$ explicitly and averaging 
over the initial conditions may produce these terms. For example, for the initial thermal 
state with the Bose distribution function and in the absence of the interactions the Kedlysh Green function,
 which is $G_K^0\sim \langle\phi_c\phi_c\rangle_{i.c.}$, contains combination (see eq.\eqref{GK0})
$$
G_K^0\sim \frac{\hbar}{\omega_p}\coth\left(\frac{\hbar\omega_p}{T}\right).
$$
In the zero temperature limit $G_K^0\sim \hbar/\omega_p$ which is linear in $\hbar$, whereas for high temperature $G_K^0\sim T/\omega_p^2$ and this contribution is pure classical and independent from $\hbar$.

\subsection{Quantum Corrections}

The quantum corrections to the CSA (or the next-to-leading order of the semiclassical expansion) can be found with the help of the second term of the expansion (eq.\eqref{semi}). The integration over $\varphi_q$ can not be performed straightforwardly because of the new $\varphi_q^3$ term.\footnote{The same problem arise during the calculation of the correlation function like $G_R(x,x')$.}
 However, each
 $\varphi_q$ can be replaced by the functional derivative over the source $J$
 due to $\varphi_q J $ term in the Keldysh action (eq.\eqref{actionJ}) as
 \begin{gather}\label{J}
  \varphi_q(x)e^{i S_K[\varphi_c,\varphi_q,J]} = -i \frac{\delta }{\delta J(x)} e^{i S_K[\varphi_c,\varphi_q,J]} .
 \end{gather}
 Then the quantum corrections to the CSA averages are \footnote{A similar $\hbar^2$ expansion
 was studied in the paper by B\"odeker \cite{Bodeker97}, where $\hbar^2$ contributions
 both from the initial state and the dynamical evolution were considered all together. 
 In the later work \cite{Bodeker98} it was argued that the dynamical $\hbar^2$ contribution 
 (analogous to eq. \eqref{NLO} ) dominates at large times. }
 \begin{gather}\label{NLO}
   \langle \hat O \rangle = \Bigg\langle O[\phi_c(x)] + 
   \frac{g\hbar^2}{4} \int dy\ \phi_c(y) \frac{\delta^3 O[\phi_c(x)]}{\delta J(y)^3} \Bigg|_{J=0}
   \Bigg\rangle_{i.c}.
 \end{gather}
 The recipe of eq.\eqref{NLO} similar to the CSA one:

- find the classical trajectory as a function of the initial
conditions;

- perform three variations over the auxiliary source (not really needed);

- integrate over intermediate time and average with the Wigner function.\\
 It is easy to recast
all terms of the semiclassical approximation to the following general form
 \begin{gather}
   \label{all_terms}
  \langle \hat O \rangle =
  \Bigg\langle\bar T e^{ \frac{g\hbar^2}{4}\displaystyle
   \int dy\
   \phi_{c}(y)
   \frac{\delta^3}{\delta J^3(y)}
   }\ O[\phi_{c}(x)]\Bigg\rangle_{i.c.}
 \end{gather}
 Here $\bar T$ denote the anti-time ordering which is required
 to recover exponential form. The eq.\eqref{all_terms} shows that
 the building block of the semiclassical expansion is the full
 nonperturbative solution of the classical EoM $\phi_{cl}(x)$ and its variations over the additional source $J(x)$. 

 It turns out that it is not necessary to calculate the variations of the classical solution explicitly. 
 Let us define n-th variation as 
 \begin{equation}\label{var}
  \Phi_n(x;x_1,x_2,\ldots  x_n)= 	\frac{\delta^n\phi_{c}(x)}{\delta J( x_1)\delta J( x_2)\ldots\delta J( x_n)}.
	\end{equation}
  $\Phi_n(x;x_1,x_2,\ldots  x_n)$ can be calculated by variation of the classical equation of motion
\begin{gather}
 \frac{\delta^n}{\delta J( x_1)\ldots\delta J( x_n)}
  \Bigg(\partial_{\mu}\partial^{\mu}\phi_{c}( x) +g\phi_{c}^3( x) =J(x)\Bigg),\label{var_eq}\\
\hat L_{\phi}\Phi_1( x;x_1) = \delta^{(4)}( x -x_1),\nonumber\\
\hat  L_{\phi}\Phi_2( x; x_1, x_2) = -6 g \phi_{c}( x)\Phi_1( x; x_1)\Phi_1( x; x_2),\nonumber\\
 \hat L_{\phi}\Phi_3(x; x_1,x_2,x_3) =
-6g \phi_c(x) \Phi_1(x;x_1)\Phi_2(x;x_2,x_3) \nonumber\\
-6g \phi_c(x) \Phi_1(x;x_2)\Phi_2(x;x_1,x_3) \nonumber\\
-6g \phi_c(x) \Phi_1(x;x_3)\Phi_2(x;x_1,x_2) \nonumber\\
-6g \Phi_1(x;x_1)\Phi_1(x;x_2)\Phi_1(x;x_3),\nonumber\\
\cdots\nonumber\\
 \hat L_{\phi} =\partial_\mu\partial^\mu +m^2 + 3 g \phi_{c}^2( x) \equiv\hat L_0 + 3 g\phi_{c}^2( x)\nonumber .
\end{gather}
Hence, to calculate the quantum correction to the CSA one need
to find the solution  of the $n$
 coupled differential equations  without knowledge
 of the exact dependence of the classical solution $\phi_{c}(x)$ from the
 auxiliary source $J(x)$. The initial conditions for these equations are zero ones because of the causality 
 ($\phi_c(x)$ depends on the source $J$ only at the preceding times).

\section{Comparison $g^2$ and $\hbar^2$ expansions}
\label{sec:comparison}

Now we are ready to compare the perturbative and the semiclassical approaches up to two loops.
For this purpose, we perform the semiclassical expansion of the $G_R(x_1,x_2)$ up to $\hbar^2$ terms and show how this result, 
being decomposed further up to $g^2$, reproduce the perturbative 
answer of the section \ref{sec:perturbative}.

Let us consider the full retarded Green function and expand it according to eq.\eqref{semi}, eq.\eqref{J}, and eq.\eqref{var}
\begin{multline}\label{GRh2}
  G_R(x_1,x_2) = -i\langle \varphi_{cl}(x_1)\varphi_q(x_2)\rangle = 
  - \left\langle \Phi_1(x_1;x_2)\right\rangle_{i.c.}\\
  + \frac{g\hbar^2}{4}
  \Big\langle\int dy \big(\Phi_1(y;x_2)\Phi_3(x_1;y,y,y) \\
  +\phi_c(y)\Phi_4(x_1;y,y,y,x_2)\big)\Big\rangle_{i.c.}.
\end{multline}

Let us denote the Leading Order retarded Green function as $G_R^{CSA}(x_1,x_2)$, then from eq.\eqref{GRh2} it is obvious that
\begin{gather}
  G_R^{CSA}(x_1,x_2) = -\left\langle \Phi_1(x_1;x_2)\right\rangle_{i.c.},\\
  \hat L_\phi G_R^{CSA}(x_1,x_2) = - \delta^{(d)}(x_1-x_2).
\end{gather}

The result of eq.\eqref{GRh2} presents the Leading and Next-to-Leading orders of the semiclassical expansion;
however, it is still the full nonperturbative answer in the sense of the coupling constant. In order to perform expansion in $g$ 
we need to express $\phi_c(x)$ and $\Phi_1(x_1;x_2)$ through the non-interacting counterparts
\begin{gather}\label{phi_ccc}
  \phi_c(x) = \phi_0(x) +g \int dy\ G_R^0(x,y)\phi_c^3(y),\\
 \Phi_1(x_1;x_2) = -G_R^0(x_1,x_2) + 3g\int dy\ G_R^0(x_1,y)\phi_c^2(y)\Phi_1(y,x_2),\nonumber
\end{gather}
where $\phi_0(x)$ and $G_R^0(x_1,x_2)$ are solutions of the free differential equation (eq.\eqref{L0})
\begin{gather}
  \hat L_0 \phi_0 = 0,\\
  \hat L_0 G_R^0(x_1,x_2) = - \delta^{(d)}(x_1-x_2).
\end{gather}
The iterative expansion of the eq.\eqref{phi_ccc} up to $g^2$ is the following
\begin{gather}
  \phi_c(x) = \phi_0(x)+ g\int dy\ G_R^0(x,y)\phi_0^3(y)\nonumber\\
   + 3g^2\int dy\ G_R^0(x,y)\phi_0^2(y)\int dz\ G_R^0(y,z)\phi_0^3(z) + O(g^3),\\
   \Phi_1(x_1;x_2) = -G_R^0(x_1,x_2) - 3g \int dy\ G_R^0(x_1,y)\phi_0^2(y)G_R^0(y,x_2)\nonumber\\
   -9g^2\int dy\ G_R^0(x_1,y)\phi_0^2(y)\int dz\ G_R^0(y,z)\phi_0^2(z)G_R^0(z,x_2)\nonumber\\
   -6g^2\int dy\ G_R^0(x_1,y)\phi_0(y)G_R^0(y,x_2)\int dz\ G_R^0(y,z)\phi_0^3(z).\nonumber\label{expan}
\end{gather}

The higher variations $\Phi_3$ and $\Phi_4$ can be rewritten through $\phi_c$ and $\Phi_1$ with the help of the integration representations 
of the differential equations of eq.\eqref{var_eq}. However, it is enough to expand the higher variation only up to  $g$, 
because of the addition power of $g$ in the second term of
 the eq.\eqref{GRh2}. Moreover, the contribution of the $\Phi_4$ vanishes, because the lowest term in this variation proportional to $g^2$.
 The remaining $h^2$ term of the eq.\eqref{GRh2} is
 \begin{multline*}
  \frac{g\hbar^2}{4}
  \int dy\ \Phi_1(y;x_2)\Phi_3(x_1;y,y,y)\to \\
  -\frac{3g^2\hbar^2}{2}\int dy\ 
  G_R^0(y,x_2) \int dz\ G_R^0(x_1,z)[G_R^0(z,y)]^3.
 \end{multline*}

 Let us draw the contributions to the full retarded Green function $G_R(x_1,x_2)$ pictorially.

 \begin{gather}\label{expan_pict}
   \begin{tikzpicture}[baseline=(a.base)]
       \begin{feynhand}
       \vertex (a) at (0,0);\vertex (b) at (2.0,0) ;
       \setlength{\feynhandlinesize}{2pt}
       \propag [fer] (a) to (b);
       \end{feynhand}
   \end{tikzpicture}=
   \begin{tikzpicture}[baseline=(a.base)]
       \begin{feynhand}
       \vertex (a) at (0,0);\vertex (b) at (2.0,0);
       \propag [fer] (a) to (b);
       \end{feynhand}
   \end{tikzpicture}+\big\langle 
   \begin{tikzpicture}[baseline=(a.base)]
       \begin{feynhand}
       \vertex (a) at (0,0);\vertex [dot] (b) at (1.5,0) {};\vertex [dot, Gray] (b1) at (1.2,0.4) {};\vertex [dot, Gray] (b2) at (1.8,0.4) {};  \vertex (c) at (3.0,0);
       \propag [fer] (a) to (b);
       \propag [fer] (b) to (c);
       \propag [plain, Gray] (b) to (b1);
       \propag [plain, Gray] (b) to (b2);
       \end{feynhand}
   \end{tikzpicture}\big\rangle_{i.c.} \nonumber \\ + 
   \big\langle
   \begin{tikzpicture}[baseline=(a.base)]
       \begin{feynhand}
       \vertex (a) at (0,0);\vertex [dot] (b) at (1.5,0) {};\vertex [dot, Gray] (b1) at (1.2,0.4) {};\vertex [dot, Gray] (b2) at (1.8,0.4) {};\vertex [dot] (c) at (3.0,0) {};\vertex [dot, Gray] (c1) at (2.7,0.4) {};\vertex [dot, Gray] (c2) at (3.3,0.4) {};\vertex (d) at (4.5,0);
       \propag [fer] (a) to (b);
       \propag [fer] (b) to (c);
       \propag [fer] (c) to (d);
       \propag [plain, Gray] (b) to (b1);
       \propag [plain, Gray] (b) to (b2);
       \propag [plain, Gray] (c) to (c1);
       \propag [plain, Gray] (c) to (c2);
       \end{feynhand}
   \end{tikzpicture}\big\rangle_{i.c.}  \nonumber\\ 
   + \big\langle
   \begin{tikzpicture}[baseline=(a.base)]
       \begin{feynhand}
       \vertex (a) at (0,0);\vertex [dot] (b) at (1.5,0) {};\vertex [dot] (bb) at (0.9,0.8) {};\vertex  [dot, Gray] (b1) at (1.8,0.4) {};\vertex  [dot, Gray] (bb1) at (0.6,1.2) {};\vertex  [dot, Gray] (bb2) at (0.9,1.3) {};\vertex  [dot, Gray] (bb3) at (1.2,1.2) {};\vertex (c) at (3.0,0);
       \propag [fer] (a) to (b);
       \propag [fer] (b) to (c);
       \propag [fer] (bb) to (b);
       \propag [plain, Gray] (bb) to (bb1);
       \propag [plain, Gray] (bb) to (bb2);
       \propag [plain, Gray] (bb) to (bb3);
       \propag [plain, Gray] (b) to (b1);
       \end{feynhand}
   \end{tikzpicture}\big\rangle_{i.c.} \nonumber\\ +
   \begin{tikzpicture}[baseline=(a.base)]
       \begin{feynhand}
       \vertex (a) at (0,0);\vertex [ringdot] (b) at (1.5,0) {};\vertex [dot] (c) at (3.0,0) {};\vertex (d) at (4.5,0);
       \propag [fer] (a) to (b);
       \propag [fer] (b) to (c);
       \propag [fer] (c) to (d);
       \propag [fer] (b) to [in=120,out=60] (c);
       \propag [fer] (b) to [in=240,out=300] (c);
       \end{feynhand}
   \end{tikzpicture}+...
 \end{gather}
 
All lines and vertices have the same meaning as in section \ref{sec:perturbative}. The only new element - the grey blob - 
denotes the free field $\phi_0(x)$. Since only $\phi_0(x)$ depends on the initial conditions in the above expansion, it is straightforward 
to perform averaging according to the rule

\begin{multline}\label{wick}
 \langle\phi_0(x)\phi_0(y)\rangle_{i.c.}=
 \big\langle
 \begin{tikzpicture}[baseline=(a.base)]
   \begin{feynhand}
   \vertex (a) at (0,0) {x};\vertex [dot, Gray] (a1) at (0.8,0) {};\vertex (b) at (2,0) {y};\vertex [dot, Gray] (b1) at (1.2,0) {};
   \propag [plain, Gray] (a) to (a1);
   \propag [plain, Gray] (b) to (b1);
   \end{feynhand}
 \end{tikzpicture}
 \big\rangle_{i.c.}= \\
 =  \begin{tikzpicture}[baseline=(a.base)]
   \begin{feynhand}
   \vertex (a) at (0,0) {x};\vertex (b) at (2.0,0) {y};
   \propag [plain] (a) to (b);
   \end{feynhand}
 \end{tikzpicture} = iG_K^0(x;y).
\end{multline}
Since we consider the Gaussian form of the Wigner function (to satisfy the demands of the perturbative approach), the eq.\eqref{wick} represents the
basic element of the Wick's theorem - the contraction of two $\phi_0(x)$. For example, the "cactus" diagram, that we mention earlier 
in eq.\eqref{cactus}, is recovered from the fourth term of the expansion (eq.\eqref{expan_pict}), or the last line of eq.\eqref{expan} .

 $$
   \big\langle
   \begin{tikzpicture}[baseline=(a.base)]
       \begin{feynhand}
       \vertex (a) at (0,0);\vertex [dot] (b) at (1.5,0) {};\vertex [dot] (bb) at (0.9,0.8) {};\vertex  [dot, Gray] (b1) at (1.8,0.4) {};\vertex  [dot, Gray] (bb1) at (0.6,1.2) {};\vertex  [dot, Gray] (bb2) at (0.9,1.3) {};\vertex  [dot, Gray] (bb3) at (1.2,1.2) {};\vertex (c) at (3.0,0);
       \propag [fer] (a) to (b);
       \propag [fer] (b) to (c);
       \propag [fer] (bb) to (b);
       \propag [plain, Gray] (bb) to (bb1);
       \propag [plain, Gray] (bb) to (bb2);
       \propag [plain, Gray] (bb) to (bb3);
       \propag [plain, Gray] (b) to (b1);
       \end{feynhand}
   \end{tikzpicture}\big\rangle_{i.c.}=3\ 
   \begin{tikzpicture}[baseline=(a.base)]
     \begin{feynhand}
     \vertex (a) at (0,0);\vertex [dot] (b) at (1.5,0) {};\vertex [dot] (bb) at (1.5,0.8) {};\vertex  (bbb) at (1.5,1.6);\vertex (c) at (3.0,0);
     \propag [fer] (a) to (b);
     \propag [fer] (b) to (c);
     \propag [fer] (bb) to [in=135,out=225] (b);
     \propag [plain] (bb) to [in=45,out=315] (b);
     \propag [plain] (bb) to [in=180,out=135] (bbb);
     \propag [plain] (bbb) to [in=45,out=0] (bb);
     \end{feynhand}
 \end{tikzpicture}
 $$
 All other contributions of the expansion \eqref{exp_g} are recovered correspondingly.

 One can see, that the Leading Order semiclassical term (the CSA) reproduce all the contributions of the $g^2$ terms of the perturbative approach 
 except the last one. However, this term is subleading for a highly occupied initial state as we discuss in the next section.

 \section{Shear viscosity and the CSA applicability}
 \label{sec:viscosity}

 In this section, we apply the semiclassical formalism to the evaluation of the transport coefficients.
 The CSA is useful for numerical calculations since the path-integrals over initial conditions can be done with the help of the Monte-Carlo approach \cite{jakovac15,bogusl18}. Another advantage of the CSA is the possibility to take into account strongly correlated initial conditions (non-Gaussian ones).
  We consider shear viscosity as an example. In order to evaluate it one can use the Kubo linear response theory \cite{KUBO,Hosoya,jeonVisc,jakovac99,jakovac15,WHZ96,WH99,CzaJeon}, where
   transport coefficients can be expressed through the retarded correlator $R^{\mu\nu}_{\alpha\beta}$ of two components of the stress-energy tensor $T^{\mu\nu}$ as
 \begin{equation}
   R^{\mu\nu}_{\alpha\beta}(x;x')=-\frac{i}{\hbar}\theta(t-t')\langle[\hat T^{\mu\nu}(x),\hat T_{\alpha\beta}(x')]\rangle.
 \end{equation}
 The Kubo theory is valid if a system is in a (quasi)stationary state when hydrodynamical 
 desctription \cite{jeonheintz} is applicable and $R^{\mu\nu}_{\alpha\beta}(x;x')$ depends only on $x-x'$. 
 In the rest frame, the shear viscosity can be expressed through the Fourier transform of the (12-12) correlation function
 \begin{equation*}
   R^{12}_{12}(p)=\int d^4(x-x') e^{ip^\mu(x_\mu-x_\mu')}R^{12}_{12}(x;x')
 \end{equation*}
 as
 \begin{equation*}
   \eta = i\lim_{p^0\to0}\lim_{p^i\to 0}\partial_0 R^{12}_{12}(p).
 \end{equation*}
 In the general frame there might be the energy flow and one can define the flow 
 velocity $u^\mu$ for energy current as the only time-like eigenvector of the average
  strees-energy tensor $\langle\hat T^{\mu\nu}\rangle$ with eigenvalue equal to energy density. We normalize it as $u_\mu u^\mu=1$. 
  In that case, one can rewrite the expression for shear viscosity in the covariant form through the retarded correlator of the
   traceless part of the stress tensor. The final expression is
 \begin{equation}
   \eta(x)=-\frac{1}{10}\Delta^{\mu\nu}_{\alpha\beta}\int d^4y\ u^\rho y_\rho R^{\alpha\beta}_{\mu\nu}(x+y;x),
 \end{equation}
 where
 \begin{gather*}
   \Delta^{\mu\nu}_{\alpha\beta}=\frac12\left(\Delta^\mu_\alpha \Delta^\nu_\beta
   +\Delta^\nu_\alpha \Delta^\mu_\beta-\frac23\Delta^{\mu\nu}\Delta_{\alpha\beta}\right),\\
   \Delta^{\mu\nu}=g^{\mu\nu}-u^{\mu}u^{\nu}. 
 \end{gather*}
 From this equation one can analyse the applicability of the hydrodynamic description. Here we have two scales.
  The first one is the scale on which energy flow velocity $u^\mu$ varies or $\Delta_{\mu\nu}^{\alpha\beta}R^{\mu\nu}_{\alpha\beta}(x+y;x)$
   changes as function of $x$. The another one is the scale on which $\Delta_{\mu\nu}^{\alpha\beta}R^{\mu\nu}_{\alpha\beta}(x+y;x)$ 
   decays as function of $y$ and on this scales the integral, entering into the shear viscosity, converges. 
   Hydrodynamics is applicable if the first scale is much larger than the second one, or, in other words,
    all the microscopic dynamics enters into  the large scale behaviour only through the number of transport coefficients. 
 
 For the system under consideration  the stress energy tensor equals to
 \begin{equation}
   T^{\mu\nu}=\partial^\mu\varphi\partial^\nu\varphi
   -g^{\mu\nu}\left(\frac12\partial^\rho\varphi\partial_\rho\varphi-
   \frac12m^2\varphi^2-\frac{g}{4}\varphi^4\right)
 \end{equation}
 and only first part $\Theta^{\mu\nu}(x)=\partial^\mu\varphi(x)\partial^\nu\varphi(x)$ contribute to the shear viscosity.
  It means that we need to evaluate retarded correlator which is proportional to
 $ \langle \Theta^{\mu\nu}_{cl}(x)\Theta^{\alpha\beta}_{q}(x')\rangle $ where the definition of "classical" and "quantum" 
 components are the same as before
 \begin{gather*}
   \Theta^{\mu\nu}_{cl}(x)=\frac12\left(\Theta^{\mu\nu}_{F}(x)+\Theta^{\mu\nu}_{B}(x)\right)\\
   \hbar\Theta^{\mu\nu}_{q}(x)=\Theta^{\mu\nu}_{F}(x)-\Theta^{\mu\nu}_{B}(x).
 \end{gather*}
 In terms of $\varphi_{cl(q)}$ the result is
 \begin{multline*}
   R(x;x')\equiv\Delta_{\mu\nu}^{\alpha\beta}R^{\mu\nu}_{\alpha\beta}(x;x')=\\=-2i\Delta_{\mu\nu}^{\alpha\beta}\langle
   \partial^\mu\varphi_{cl}(x)\partial^\nu\varphi_{cl}(x)\partial_\alpha'\varphi_{cl}(x')\partial_\beta'\varphi_q(x')\rangle,
 \end{multline*}
 where $\partial'_\alpha = \frac{\partial}{\partial x_\alpha'}$. 
 Now we can apply the semiclassical approach and derive the leading order contribution for $R(x;x')$. The result is
 \begin{gather}
   R(x;x')=-4\Delta_{\mu\nu}^{\alpha\beta}\langle\partial^\mu\phi_c(x)\partial'_{\alpha}\phi_c(x')\partial^\nu\partial_\beta'\Phi_1(x;x')\rangle_{i.c.}.
   \label{viscres}
 \end{gather}
 In principle, by taking variational derivatives one can obtain the next-to-leading order corrections.
  However, even the leading order is nonperturbative in coupling constant and contains a lot of diagrams. 
  Let us explicitelly consider two similar diagrams of eq.\eqref{shear}.  The first diagram contributes to the leading order shear viscosity 
  and it is taken into account by the CSA, whereas the second one proportional to $\hbar^2$ and belongs to the NLO semiclassical term. 
 
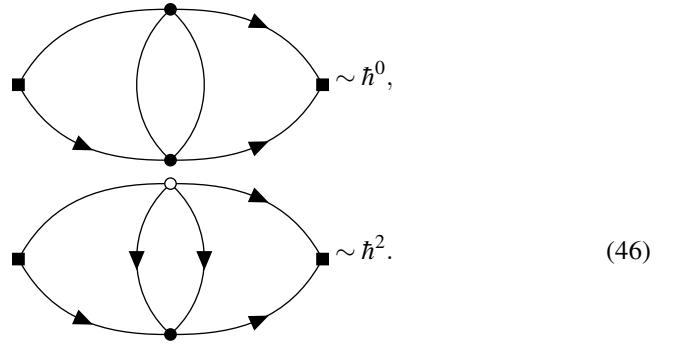
\begin{gather}\label{shear}
 \begin{tikzpicture}[baseline=(a.base)]
   \begin{feynhand}
   \vertex (a) [squaredot] at (0,0) {};\vertex (b) [squaredot] at (4.0,0) {};\vertex [dot] (c) at (2,1) {};\vertex [dot] (d) at (2,-1) {};
   \propag [plain] (a) to [in=180,out=60] (c);
   \propag [fer] (c) to [in=120,out=0](b);
   \propag [fer] (a) to [in=180,out=300](d);
   \propag [fer] (d) to [in=240,out=0](b);
   \propag [plain] (c) to [in=135,out=225](d);
   \propag [plain] (c) to [in=45,out=315](d);
   \end{feynhand}
 \end{tikzpicture}\sim \hbar^0,\nonumber\\
 \begin{tikzpicture}[baseline=(a.base)]
   \begin{feynhand}
   \vertex (a) [squaredot] at (0,0) {};\vertex (b) [squaredot] at (4.0,0) {};\vertex [ringdot] (c) at (2,1) {};\vertex [dot] (d) at (2,-1) {};
   \propag [plain] (a) to [in=180,out=60] (c);
   \propag [fer] (c) to [in=120,out=0](b);
   \propag [fer] (a) to [in=180,out=300](d);
   \propag [fer] (d) to [in=240,out=0](b);
   \propag [fer] (c) to [in=135,out=225](d);
   \propag [fer] (c) to [in=45,out=315](d);
   \end{feynhand}
 \end{tikzpicture}\sim \hbar^2.
\end{gather}
One can observe that the difference comes only from the central loop. In the first case, it contains 
the product of two Keldysh Green functions $\sim G_K^0G_K^0$, whereas the second one has 
$\sim G_R^0G_R^0$ insertion. Every Keldysh Green function has $2f_p+1$ multiplier in contrast to the
 retarded one. If the initial state is highly occupied, then $f_p\gg 1$, and we can neglect the second contribution. 
 This analysis can be extended to any diagram, contributing to the viscosity or any other observable. 
 For each diagram $\sim \hbar^{2n}$ there is the diagram $\sim \hbar^0$ which differs by $2n$ times
  substitutions of $G_R^0$ by $G_K^0$. In other words, in this diagram $n$ "white" vertices are changed 
  to the "black" ones. The resulting diagram is greater due to $2f_p+1$ factors and is already included in the CSA.
   That explains why the CSA works well for the highly excited initial state and sums up all leading contributions. 
    Hence, the results of works \cite{aarts98,aarts01} are clarified.

 \section{$\hbar^2$ diagram technique}
\label{sec:newdiag}
 
 In order to systematically improve the CSA and analyze the higher order corrections 
 in this section we present alternative diagram technique which can be used to construct all the diagrams at given $\hbar^2$ order. Let us start again from the general expression for the observable  (eq.\eqref{common}) and shift integration variable $\varphi_{cl}(x)=\phi_c(x)+\tilde\varphi_{cl}(x)$, where $\phi_c(x)$ is again the solution of the classical equation of motion with the corresponding boundary conditions. It means that $\tilde\varphi_{cl}$ obeys zero boundary conditions and all 
 dependence on $\alpha(\bx)$ and $\Pi(\bx)$ enters only through $\phi_c(x)$. Then the full retarded Green function can be written as
 \begin{multline}
   G_R(x;x')=-i\Big\langle\int\mathcal D\tilde\varphi_{cl}\int\mathcal D\varphi_q\ \tilde\varphi_{cl}(x)\varphi_{q}(x')\\
  \times e^{-i\int d^dx\big(\varphi_q(x)\hat L_\phi\tilde\varphi_{cl}(x)+
  3g\phi_c(x)\tilde\varphi^2_{cl}(x)\varphi_q(x)+g\tilde\varphi^3_{cl}(x)\varphi_q(x)\big)}\\
   \times e^{-i\frac{g\hbar^2}{4} \int d^dx\big(\phi_{c}(x)\varphi_q^3(x)+\varphi_{cl}(x)\varphi_q^3(x)\big)}\Big\rangle_{i.c.}.
 \end{multline}
 Now we perform the perturbative expansion of the above expression (without averaging over the initial conditions yet).
  One can check that due to zero initial conditions for $\tilde\varphi_{cl}$ there is only one non-zero contraction
 \begin{equation}
   \langle\langle \tilde\varphi_{cl}(x)\varphi_{q}(x')\rangle\rangle=-i\Phi_1(x;x')
 \end{equation}

 where by $\langle\langle\ldots \rangle\rangle$ we denote the functional integration without averaging over the initial conditions. This contraction is represented by the dashed line in diagrams
 $$
   \begin{tikzpicture}[baseline=(a.base)]
     \begin{feynhand}
       \vertex (a) at (0,0) {$x'$};\vertex (b) at (3.0,0) {$x$};
       \propag [chasca] (a) to (b);
     \end{feynhand}
   \end{tikzpicture}\quad -i\Phi_1(x;x').
 $$
 Also we have four vertices in the theory
 $$\begin{tikzpicture}[baseline=(v.base)]
     \begin{feynhand}
       \vertex [dot] (v) at (0,0) {};\vertex (c1) at (-0.8,0.5) {};\vertex (c2) at (-1,0) {};
       \vertex (c3) at (-0.8,-0.5) {};\vertex (q) at (1.0,0.0) {};
       \propag [sca] (c1) to (v);
       \propag [sca] (c2) to (v);
       \propag [sca] (c3) to (v);
       \propag [chasca] (v) to (q);
     \end{feynhand}
   \end{tikzpicture}\ -ig,\quad
   \begin{tikzpicture}[baseline=(v.base)]
     \begin{feynhand}
       \vertex [ringdot] (v) at (0,0) {};\vertex (q1) at (0.8,0.5) {};\vertex (q2) at (1,0) {};
       \vertex (q3) at (0.8,-0.5) {};\vertex (c) at (-1.0,0.0) {};
       \propag [chasca] (v) to (q1);
       \propag [chasca] (v) to (q2);
       \propag [chasca] (v) to (q3);
       \propag [sca] (c) to (v);
     \end{feynhand}
   \end{tikzpicture}\ -\frac{ig\hbar^2}{4},
 $$
 $$
 \begin{tikzpicture}[baseline=(v.base)]
     \begin{feynhand}
       \vertex [dot] (v) at (0,0) {};\vertex (c1) at (-0.8,0.5) {};\vertex [crossdot] (c2) at (-0.6,0) {};
       \vertex (c3) at (-0.8,-0.5) {};\vertex (q) at (1.0,0.0) {};
       \propag [sca] (c1) to (v);
       \propag [plain] (c2) to (v);
       \propag [sca] (c3) to (v);
       \propag [chasca] (v) to (q);
     \end{feynhand}
   \end{tikzpicture}\ -3ig\phi_c(x),\quad
   \begin{tikzpicture}[baseline=(v.base)]
     \begin{feynhand}
       \vertex [ringdot] (v) at (0,0) {};\vertex (q1) at (0.8,0.5) {};\vertex (q2) at (1,0) {};
       \vertex (q3) at (0.8,-0.5) {};\vertex [crossdot] (c) at (-0.6,0.0) {};
       \propag [chasca] (v) to (q1);
       \propag [chasca] (v) to (q2);
       \propag [chasca] (v) to (q3);
       \propag [plain] (c) to (v);
     \end{feynhand}
   \end{tikzpicture}\ -\frac{ig\hbar^2\phi_c(x)}{4}.
 $$
 After that, the perturbative expansion can be performed in a standard manner.
  However, this expansion has an important property which is absent in the standard Keldysh-Schwinger technique considered before.
   $\Phi_1(x;x')$ is nonzero only for $t>t'$ due to causality.
    It means that any loop of the form $\Phi_1(x;x_1)\Phi_1(x_1;x_2)...\Phi_1(x_n,x)$ is zero and one can fix the 
    direction of time flow on each diagram. Suppose we want to consider all diagrams that correspond to $\hbar^{2n}$ 
    order for some $n$. The number of diagrams of this order is always finite, and all these diagrams have $2n$-loop.
     The proof is based on the observation that the "black" vertices increase the number of lines in time, 
     whereas the "white" vertices decrease. But the number of the "white" vertices should be $n$, so we have the only finite number of possibilities to insert the "black" vertices. In the standard approach, the number of the diagrams for a given power of $\hbar$ is infinite. So this approach effectively performs resummation of many diagrams and combines it into the finite number. 

Also we never use here the precise form of the initial Wigner function, and it is not necessary Gaussian. Hence, with the help of this technique, we can treat correlated initial state. All complexity of the initial state comes only on the last step - averaging over the initial condition which enters through $\phi_c(x)$.

For example, the expression for the viscosity \eqref{viscres} can be represented in this technique by the only diagram
$$
R(x,x')=\Big\langle
\begin{tikzpicture}[baseline=(a.base)]
  \begin{feynhand}
    \vertex [squaredot] (a) at (0,0) {};\vertex [squaredot] (b) at (4.0,0) {};
    \vertex [crossdot] (a1) at (0.3,0.5) {};\vertex [crossdot] (b1) at (3.7,0.5) {};
    \propag [chasca] (a) to (b);
    \propag [plain] (a) to (a1);
    \propag [plain] (b) to (b1);
  \end{feynhand}
\end{tikzpicture}\Big\rangle_{i.c.},
$$
which is after the averaging over initial conditions is corresponded to the infinite sum of multiloop diagrams of the stanmdard Keldysh-Schwinger technique
\begin{multline}
\big\langle
\begin{tikzpicture}[baseline=(a.base)]
  \begin{feynhand}
    \vertex [squaredot] (a) at (0,0) {};\vertex [squaredot] (b) at (2.2,0) {};
    \vertex [crossdot] (a1) at (0.2,0.3) {};\vertex [crossdot] (b1) at (2.0,0.3) {};
    \propag [chasca] (a) to (b);
    \propag [plain] (a) to (a1);
    \propag [plain] (b) to (b1);
  \end{feynhand}
\end{tikzpicture}\big\rangle_{i.c.}=
\begin{tikzpicture}[baseline=(a.base)]
  \begin{feynhand}
  \vertex (a) [squaredot] at (0,0) {};\vertex (b) [squaredot] at (1.8,0) {};
  \propag [plain] (a) to [in=120,out=60] (b);
  \propag [fer] (a) to [in=240,out=300](b);
  \end{feynhand}
\end{tikzpicture}
\\+
\begin{tikzpicture}[baseline=(a.base)]
  \begin{feynhand}
  \vertex (a) [squaredot] at (0,0) {};\vertex (b) [squaredot] at (2.2,0) {};\vertex [dot] (c) at (1.1,0.6) {};\vertex [dot] (d) at (1.1,-0.6) {};
  \propag [plain] (a) to [in=180,out=60] (c);
  \propag [fer] (c) to [in=120,out=0](b);
  \propag [fer] (a) to [in=180,out=300](d);
  \propag [fer] (d) to [in=240,out=0](b);
  \propag [plain] (c) to [in=135,out=225](d);
  \propag [plain] (c) to [in=45,out=315](d);
  \end{feynhand}
\end{tikzpicture}
+ 
\begin{tikzpicture}[baseline=(a.base)]
  \begin{feynhand}
  \vertex (a) [squaredot] at (0,0) {};\vertex (b) [squaredot] at (2.8,0) {};\vertex [dot] (c) at (0.9,0.6) {};\vertex [dot] (d) at (1.4,-0.6) {};\vertex [dot] (c1) at (1.9,0.6) {};
  \propag [plain] (a) to [in=180,out=60] (c);
  \propag [fer] (c1) to [in=120,out=0](b);
  \propag [fer] (a) to [in=180,out=300](d);
  \propag [fer] (d) to [in=240,out=0](b);
  \propag [fer] (c) to (d);
  \propag [plain] (c1) to (d);
  \propag [plain] (c) to [in=135,out=45](c1);
  \propag [plain] (c) to [in=225,out=315](c1);
  \end{feynhand}
\end{tikzpicture}+...
\end{multline}

 Having in hand diagrammatic technique, one can exploit a variety of methods for its resummation,
 like the summation of one-particle irreducible diagrams or elimination of tadpole contribution which may 
 produce nonperturbative both in $g$ and $\hbar$ contributions. This feature of the theory will be discussed 
 in details in future publications.
 
 Another critical question concerns the renormalization \cite{EGW2014}
 of this theory which should be reconsidered. Since the main object of the technique is not the Green
  function but the quantity which becomes retarded Green function only after averaging, its IR and UV 
behaviour might be different from the free one. It will also be discussed in another work.

\section{Conclusions}
\label{sec:conclusions}

In this work, we compare two approaches to descriptions of the nonequilibrium quantum scalar fields:\\
- The standard Keldysh-Schwinger diagram technique, which requires the Gaussian initial conditions and the small coupling constant;\\
- The semiclassical expansion, which works with the arbitrary coupling constant, but valid for 
highly excited (or highly occupied) initial states only.\\
We analyse these two expansions in the limit where the coupling constant $g$ and the Plank constant $\hbar$ are small simultaneously. We prove the consistency of these approaches and explicitly demonstrate that 
already the first term of the semiclassical expansion (the Classical Statistical Approximation) includes almost all two loop-diagrams
 of the standard perturbative approach. We show that the only remaining $g^2\hbar^2$
 diagram is small if the initial conditions are overoccupied i.e. 
the one-particle distribution function $f_p \gg 1$. On practice, this condition defines the applicability of the CSA. 

As an example of the usefulness of the semiclassical approach, we evaluate the shear viscosity in a more general case
 of nonzero energy flow. 

Also, we present a new diagram technique that combines both the advantages of the semiclassical and the Keldysh-Schwinger diagrammatic approaches. We believe that this technique allows to perform the resummation of the next-to-leading order semiclassical contributions and improve the CSA.

\begin{acknowledgements}
This work was supported by the RFBR project 18-02-40131. 
\end{acknowledgements}

\bibliography{biblio}

\end{document}